On relation between mid-latitude ionospheric ionization and quasi-trapped energetic electrons during 15 December 2006 magnetic storm

A.V. Suvorova[a,b*], L.-C. Tsai[a], A.V. Dmitriev[c,a]

[a] *Center for Space and Remote Sensing Research, National Central University, Jhongli, Taiwan*

[b] *Skobeltsyn Institute of Nuclear Physics Moscow State University, Moscow, Russia*

[c] *Institute of Space Science, National Central University, Jhongli, Taiwan*

---

**Abstract**

We report simultaneous observations of intense fluxes of quasi-trapped energetic electrons and substantial enhancements of ionospheric electron concentration (EC) at low and middle latitudes over Pacific region during geomagnetic storm on 15 December 2006. Electrons with energy of tens of keV were measured at altitude of ~800 to 900 km by POES and DMSP satellites. Experimental data from COSMIC/FS3 satellites and global network of ground-based GPS receivers were used to determine height profiles of EC and vertical total EC, respectively. A good spatial and temporal correlation between the electron fluxes and EC enhancements was found. This fact allows us to suggest that the quasi-trapped energetic electrons can be an important source of ionospheric ionization at middle latitudes during magnetic storms.

---

## 1. Introduction

Storm-time dynamics of total electron content (TEC), the integral with height of the ionospheric electron density profile, was studied comprehensively from auroral to equatorial latitudes for many decades (e.g. Mendillo, 2006). Nowadays, numerous studies concern with an unresolved problem of TEC enhancements, so-called positive ionospheric storms (e.g., Mendillo et al., 2010; Balan et al., 2010; Wei et al., 2011). Complex mechanism of thermosphere-ionosphere system response to a geomagnetic storm involves a number of different agents such as disturbance electric fields, changes in neutral winds system and neutral chemical composition, gravity waves and diffusion. However, it is very difficult to pick out the agents forming the positive ionospheric storms.

Because of its high conductivity, the ionosphere responses quickly to variations of electric field due to such effects as magnetospheric convection, ionospheric dynamo disturbance and various kinds of wave disturbances (e.g., Biktash, 2004). Balan et al. (2011) discuss an importance of thermospheric storms developing simultaneously with ionospheric ones. Relative role of prompt penetrating (under-shielded) electric field (PPEF) and equatorward neutral winds as two sources of positive ionospheric storms at low and middle latitudes is intensively discussed in literature (Lei et al., 2008; Pedatella et al., 2009; Mendillo et al., 2010; Balan et al., 2010). Possible errors in the models are attributed to under-representation of conditions at lower altitudes, variability of the neutral wind and/or coupling with auroral sources.

Recent studies and models of the ionospheric disturbances observed during a strong geomagnetic storm on 14-15 December 2006 have revealed that the PPEF and equatorward neutral winds alone can not explain a long-lasting intense positive ionospheric storm occurred over Pacific sector during maximum and recovery phase of the magnetic storm (Lei et al., 2008; Pedatella et al., 2009). Lei et al. (2008) showed that the CMIT model simulations were able to capture the positive storm effect at equatorial ionization anomaly (EIA) crest regions during 00-03 UT on December 15. On the other hand, the authors pointed out that the model was unable to reproduce the positive effects observed for several hours after 03 UT. In addition, the CMIT simulation predicted a depletion of plasma densities over the low-latitude region at 0000 to 0300 UT on 15 December that is inconsistent with the observations. Pedatella et al. (2009) reported that during this time the height of F-layer peak increased by greater than 100 km. It was assumed that the TEC increases, observed in the topside ionosphere/plasmasphere at middle to high latitudes, might be explained by effects of particle precipitation. However, experimental evidence of this assumption was not reported.

Here we analyze fluxes of energetic electrons observed by low-altitude (heights ~ 800 to 900 km) satellites of POES and DMSP fleets during magnetic storm on 15 December 2006. We demonstrate a good correlation of the mid-latitude ionospheric ionization enhancements observed over Pacific region with the intense fluxes of quasi-trapped energetic electrons.

## 2. Positive ionospheric storm

A geomagnetic storm started at about 14 UT on 14 December 2006, when a CME-driven interplanetary shock (IS) affected the Earth's magnetosphere. The storm initial phase was lasting until ~2330 UT. After that the CME-related main phase of severe geomagnetic storm began. The storm maximum with *Dst* ~ -150 nT and Kp ~ 8+ was observed after midnight of 15 December. The recovery phase started at ~ 08 UT on 15 December. The storm main and recovery phases were accompanied by a long-lasting (from 00 to 14 UT) and widely expanded (from 12 to 24 LT) strong positive ionospheric storm with the ionization enhanced up to 50 TECU (1TECU = $10^{12}$ electrons/cm$^2$) over Pacific and American regions (from 120° to 300° longitude) (Lei et al., 2008; Pedatella et al., 2009).

One of very important factors in the study of storm-time disturbances is a consistent choice of quiet-time period. Previous studies of this event used moderately disturbed day on December 13 as a day of “quiet conditions”. We use a day on December 3 when the solar and geomagnetic activity was very quiet. This choice allows revealing prominent positive ionospheric storms on the initial, main and recovery phases of the geomagnetic storm. Figure 1 demonstrates development of strong enhancement of vertical TEC (VTEC) at 00 to 06 UT on 15 December. Global ionospheric maps (GIM) of VTEC are provided every 2 hours by a world-wide network of ground-based GPS receivers. The residual VTEC (dVTEC) was calculated as a difference between the disturbed and quiet days.

The positive storms in VTEC tend to occur in the postnoon and dusk sectors above Pacific and American region. We can distinguish two branches of the VTEC enhancements at low (~10° to 20° deg) and middle (~30° to 40°) latitudes. The low-latitude positive storm is oriented strictly along the geomagnetic equator at geomagnetic latitudes of ~15°. This storm is mostly pronounced and can be explained in the frame of a continuous complex effect of daytime eastward PPEF and equatorward neutral wind (Balan et al., 2010). The positive storm at middle latitudes persists within first 6 hours and then diminishes fast after ~06 UT. It seems that the maximum of mid-latitude storm is slightly moving pole-ward from ~30° to 40° of geomagnetic latitude. There is no clear explanation of this positive storm.

Vertical profiles of electron concentration (EC) were measured in COSMIC/FS3 space-borne experiment. The EC is expressed as a number of electrons per cubic centimeter (cc). Six satellites of the COSMIC/FS3 mission produce a sounding of the ionosphere on the base of radio occultation (RO) technique, which makes use of radio signals transmitted by the GPS satellites (Hajj et al., 2000). Usually over 2500 soundings per day provide EC height profiles over ocean and land. A 3-D EC distribution is deduced through relaxation using red-black smoothing on numerous EC height profiles. This 3-D EC image is used as an initial guess to start the iterative Multiplicative Algebraic Reconstruction Technique (MART) algorithm, and 3-D tomography of the EC is then produced around whole globe with a time step of 2 hours and spatial grid of 5° in longitude, 1° in latitude, and 5 km in height (Tsai et al., 2006).

Figure 2a represents a geographic map of residual total electron content (TEC) at 04 to 06 UT on 15 December 2006. The TEC is calculated as a height integral of EC provided by the COSMIC/FS3 3-D ionospheric tomography in the range of altitudes below 830 km. Similarly to the GIM dVTEC, the residual TEC is derived by subtraction of the storm-time TEC on 15 December 2006 from the quiet-day TEC on 3 December and expressed in TECU. Comparing Figures 1 and 2a, one can see a good agreement between the spatial distribution of GIM dVTEC and residual TEC obtained at 04 to 06 UT on 15 December 2006. Note that the magnitudes of residual TEC are slightly smaller than those of GIM dVTEC, probably because of the TEC calculation is limited by the height of 830 km.

Figure 2b shows a meridional cut of EC obtained from COSMIC/FS3 3-D ionospheric tomography at 04 to 06 UT on 15 December in longitudinal range of 130° to 135°, which is covered well by the measurements. One can see a prominent low-latitude (-20° to 20°) enhancement of EC peaked at 250 to 300 km. The EC also increases at middle latitudes of ~30° to 40° in both southern and northern hemispheres. In the southern hemisphere, the maximum of mid-latitude enhancement is located at height of ~400 km.

It is important to point out that the EC enhancements expand significantly to higher altitudes (up to 600 km and above). Note that similar pattern is revealed at other longitudes above Pacific region during whole of the main phase and maximum of the geomagnetic storm from 00 to 06 UT. Elevation of the EC to higher altitudes in the equatorial region proves the presence of strong dawn-dusk electric field operating together with the equatorward neutral winds from the higher latitudes (Balan et al., 2010). At middle latitudes, the presence of elevated and widely expanded EC enhancement might indicate to operation of a magnetospheric mechanism of charged particle contribution to redundant ionization of the mid-latitude ionosphere.

## 3. Quasi-trapped electrons

Figure 3 demonstrates geographic distribution of >30 keV electron fluxes observed during magnetically quiet interval and during magnetic storm on 14-15 December 2006. The electrons are measured at altitude of 800 km by a fleet of 5 POES satellites (Huston and Pfitzer, 1998; Evans and Greer, 2004). During magnetic quiet (Figure 3a), the vast majority of electron population at low altitudes is trapped in the inner radiation belt (IRB). Because of tilted and shifted geomagnetic dipole, the lower edge of IRB sinks to the ionospheric altitudes in the region of South Atlantic Anomaly (SAA) located in the range of longitudes from -120° to 0° and latitudes from -50° to 10°. The fluxes of energetic electrons in the quiet-time SAA are moderate ($<10^5$ ($cm^{-2}$ $s^{-1}$ $sr^{-1}$)).

During magnetic storms, the electrons precipitate intensively in a wide longitudinal range from the outer and inner radiation belts to high and to middle latitudes, respectively. As one can see in Figure 3b, the storm-time fluxes of >30 keV electrons at low and middle latitudes enhance by more than 5 orders of magnitude and exceed $10^6$ particles per ($cm^2$ s sr) that might be interpreted as “equatorial aurora”. We have to point out that very intense electron fluxes are observed in the forbidden range of drift shells above Pacific region. Particles, which penetrate in this region are quasi-trapped, because they can not close the circle of azimuthal drift path around the Earth, but they are inevitably lost in the SAA region. These quasi-trapped particles can produce an additional ionization of the ionosphere, especially at high **altitudes** where the recombination rate is very low because of very rarefied atmosphere.

Figure 4 demonstrates temporal dynamics of the quasi-trapped electron fluxes and the strength of ionospheric storms together with variations of geomagnetic indices and interplanetary electric field. We compare maxima of electron fluxes (Figure 3b) with maxima of dVTEC (Figure 1). One can clearly see that the intense particle fluxes and positive ionospheric storms appear during maximum of the geomagnetic storm, which is accompanied by very large interplanetary electric field $E$y of ~5 to 10 mV/m and strong auroral activity with AE varying from 1000 to 2000 nT. The electron fluxes are more intense at longitudes of ~180° than those at longitudes of ~120°. The intense fluxes of quasi-trapped electrons coexist and correlate with the positive ionospheric storm observed at middle latitudes. The low-latitude storm has much longer duration and its maximum occurs later.

## 4. Discussion and summary

We have demonstrated that the storm-time ionospheric disturbances on 15 December 2006 exhibit two positive storms occurred at low and middle latitudes. The low-latitude positive storm in the IEA crest regions can result from continuous effects of long-lasting daytime eastward PPEF and equatorward neutral wind (Balan et al., 2010). Pedatella et al. (2009) mentioned a positive ionospheric storm observed during that time in the southern hemisphere at geographic latitudes near 50°S above Pacific. This storm was explained by effects of soft particle precipitation associated with an equatorward movement of the poleward boundary of the trough region. However, the origin of positive ionospheric storm at latitudes of ~30° to 40° is still unclear.

Here we consider the fluxes of quasi-trapped electrons at low latitudes as a possible source of the mid-latitude ionospheric storm. Using POES data on electron fluxes in energy ranges >30 keV, >100 keV and >300 keV, we find that the integral fluxes of electrons with pitch angles about 90° have very steep spectrum, which can be fitted by a power low $F$(>E, keV) =

$4.7*10^{13}$ $E^{-4.8}$ $(cm^2\ s\ sr)^{-1}$. Note that POES also measures fluxes of electrons in the loss cone (pitch angles about 0°). Those fluxes are several-order weaker than quasi-trapped ones. From the spectrum, we calculate the electron integral energy flux of JE~$1.8*10^{12}$ eV/($cm^2$ s). Using DMSP data on soft electrons in energy range below 30 keV, we find local enhancements of >1 keV electrons with integral energy fluxes up to JE~$10^{12}$ eV/($cm^2$ s). Hence, the total integral energy flux of electrons can be estimated to be ~$2.8*10^{12}$ eV/($cm^2$ s) that is equivalent to $4.5*10^{-3}$ W $m^{-2}$. Note that this energy flux is comparable to that produced by X-class strong solar flares, which ionospheric impact can achieve ~20 TECU (e.g. Tsurutani et al., 2005). In the ionosphere, enriched by oxygen with first ionization potential of 13.6 eV, this integral energy flux produces $2.1*10^{11}$ ion-electron pairs per ($cm^2$ s). In the topside ionosphere, the recombination rate of electrons decreases fast with atmospheric density and can be estimated to be ~$10^{-2}$ $s^{-1}$. Hence, the total electron content produced by the quasi-trapped electrons can be estimated to be ~2.1 $10^{13}$ ($cm^{-2}$), i.e. ~20 TECU.

Further, we have to estimate the spatial region where the electrons lose their energy in ionization of the atmospheric atoms. A precipitating electron with energy of ~30 keV and zero pitch angle is able to reach altitudes of ~90 km (e.g. Dmitriev et al., 2010). However, from POES observations we find that vast majority of the electrons is quasi-trapped and has pitch angles close to 90°. Such electrons are bouncing along the magnetic field lines about top points. This bouncing motion is very fast and has a period of a portion of second. Because of asymmetrical orientation of the geomagnetic dipole, the height of top points varies with longitude. Figure 5 shows longitudinal variation of the height of drift shells (L-shells) calculated for the geomagnetic equator using IGRF model of epoch 2005. Participating in a gradient drift, energetic electrons move eastward along the drift shells. One can see that the L-shells are descending starting from the region of Indochina at longitudes of ~120°. In this region, the height of ~900 km corresponds to the L ~ 1.05. Above the Pacific region, the altitude of top points for bouncing electron decreases with increasing longitude. The drift shells rich minimal heights in the SAA region, where practically all the particles quasi-trapped at L < 1.1 are lost. At altitudes of 1000 km and below, the period of the azimuthal drift for 30 keV electrons is ~20 hours (Lyons and Williams, 1984). Hence, the electrons with large pitch angles can make many thousands of bounces before they are lost in the SAA region.

Taking into account the specific ionization of electrons **(the energy loss per unit distance)** and standard vertical profile of the upper atmosphere **(e.g. Dmitriev et al., 2008)**, we can calculate the number of bounces between the top point at 800 km and mirror points at height *H*min until the ~30 keV electron has lost whole of the energy in ionization. By this way, we find that for *H*min below 600 km, the ~30 keV electrons lost whole energy within ~2 hours. During this time, the electrons pass eastward no more than ~30° in longitudes, because of very slow azimuthal drift. Hence, the quasi-trapped electrons, observed westward from the longitude of -150°, have a quite high chance to lose whole of their energy in ionization of the ionosphere. Because of arched magnetic field configuration, this ionization is released in the region of geomagnetic latitudes above ~20°. It is important to note that the charged particle spend most of time in vicinity of mirror point. Hence, the quasi-trapped electrons lose most of energy rather at low to middle latitudes than at the equator. That corresponds well to the spatial location of mid-latitude positive ionospheric storm.

Another important issue is conditions required for the downward transport of electrons from the IRB to the heights below 1000 km. The well-know mechanism is a radial diffusion across the drift shells. However, in the strong magnetic field at low altitudes this diffusion is very slow that results in very weak fluxes of energetic electrons at the forbidden drift shells at low latitudes. In contrast, geomagnetic storms are accompanied by a very strong penetrated electric field of dawn-dusk direction. In the nightside, this electric field is pointed westward that results in fast (a few hours) *E*x*B* drift of particles across the magnetic field lines toward the Earth. Then the electrons drift eastward through morning sector toward noon.

We can summarize that during magnetic storm, the energetic electrons (~30 keV) drift fast radially from the IRB to the ionospheric altitudes in the nightside sector. Drifting azimuthally eastward, the quasi-trapped electrons loss the energy in ionization of the atmospheric gases and, thus, produce abundant ionization of the mid-latitude ionosphere.

**Acknowledgements**

The authors thank a team of NOAA's Polar Orbiting Environmental Satellites for providing experimental data about energetic particles and Kyoto World Data Center for Geomagnetism (http://wdc.kugi.kyoto-u.ac.jp/index.html) for providing the Dst, Kp and AE geomagnetic indices. The Global Ionosphere Maps/VTEC Data from European Data Center was obtained through ftp://ftp.unibe.ch/aiub/CODE/. The ACE solar wind data were provided by N. Ness and D. J. McComas through the CDAWeb web site. We acknowledge Dr. Patrick Newell for help with providing DMSP data. The DMSP particle detectors were designed by Dave Hardy of AFRL, and data obtained from JHU/APL. This work was supported by grant NSC 99-2811-M-008-093 and Ministry of Education under the Aim for Top University program at NCU of Taiwan #985603-20.

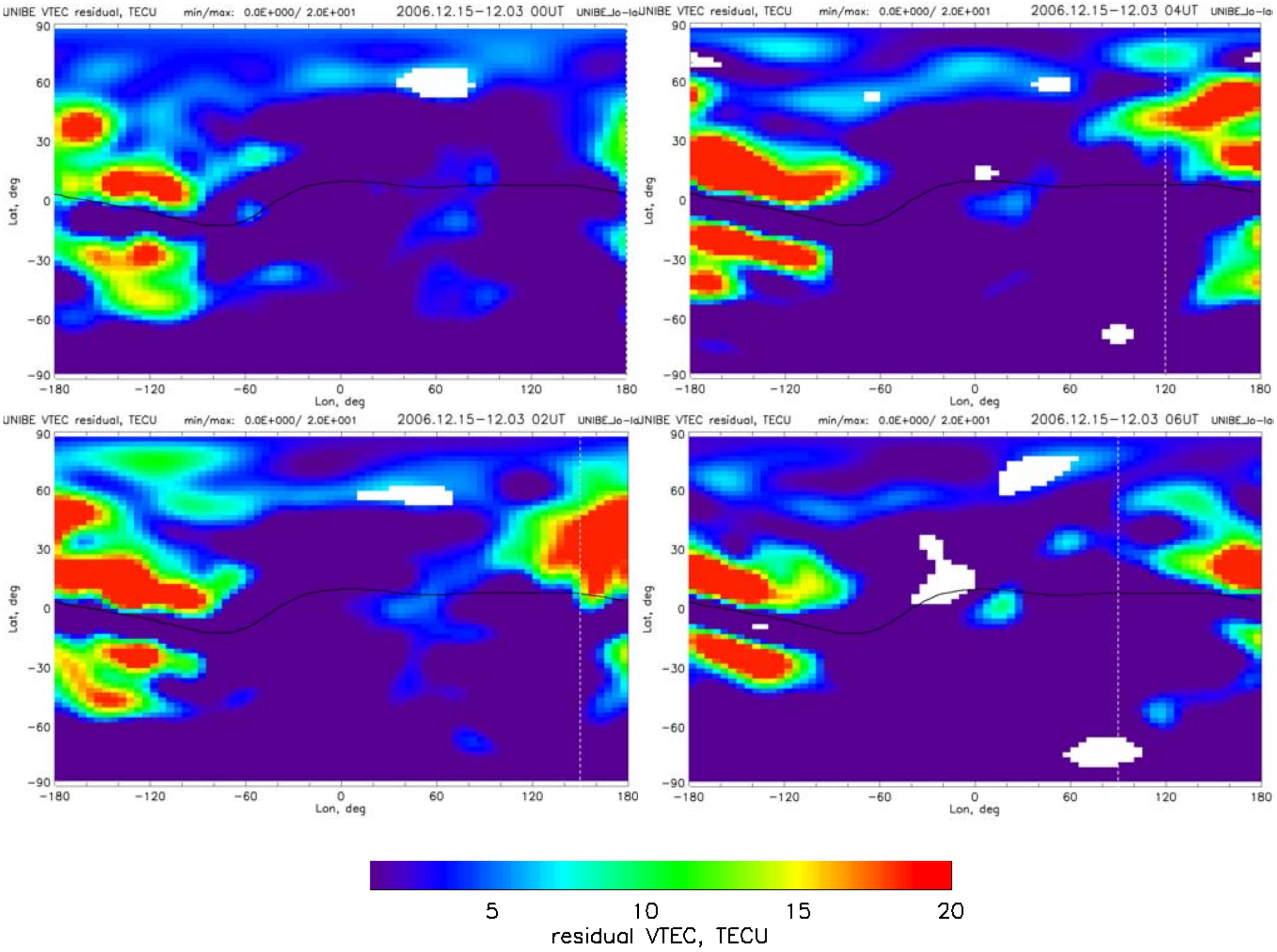

Figure 1: Global ionospheric maps of residual vertical total electron content (VTEC) between the quiet day on December 3 and disturbed days on 15 December 2006 at 00-06 UT. Geomagnetic equator is indicated by black curve. Local noon is depicted by vertical white dashed line. Strong positive ionospheric storms are visible as large red spots.

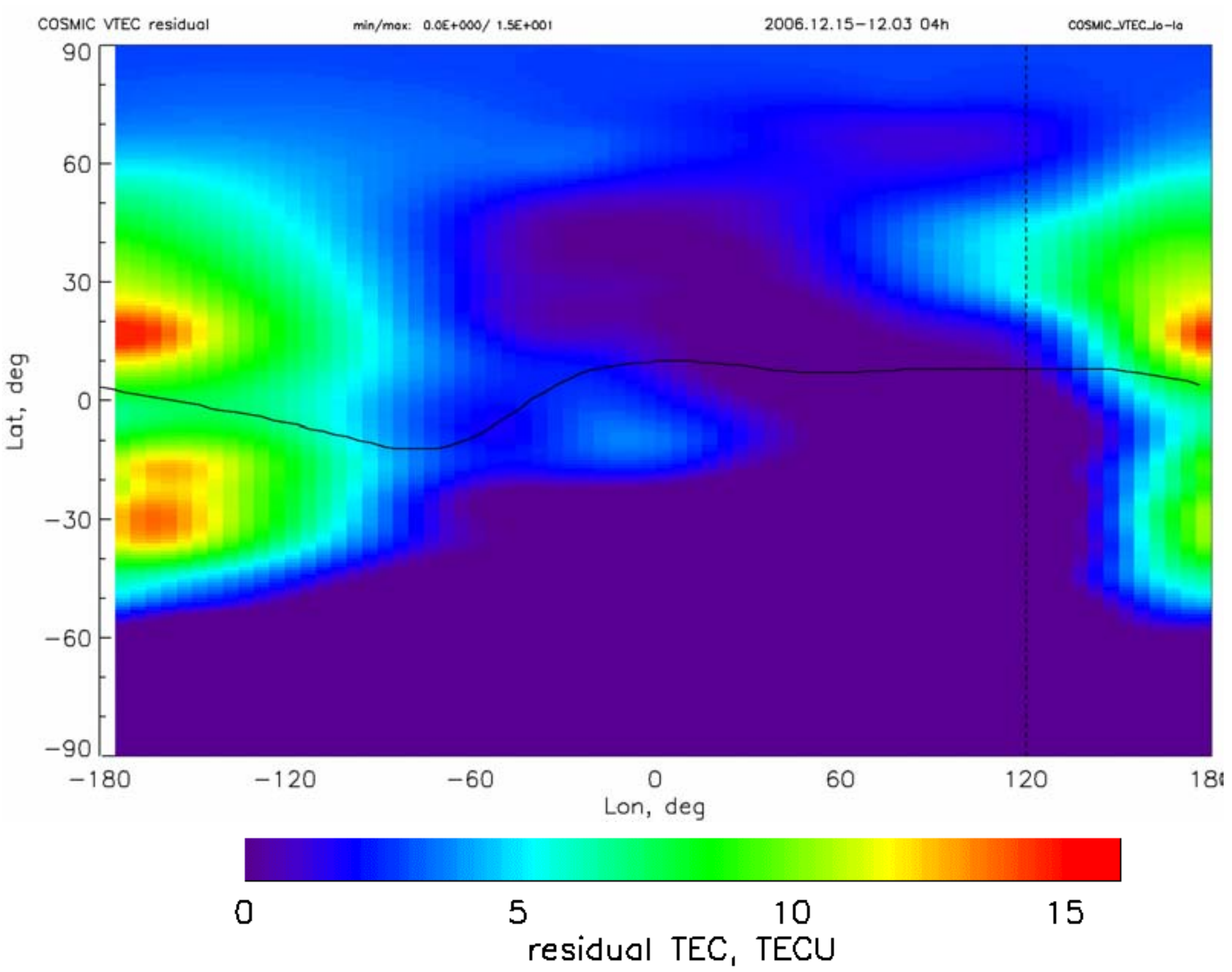

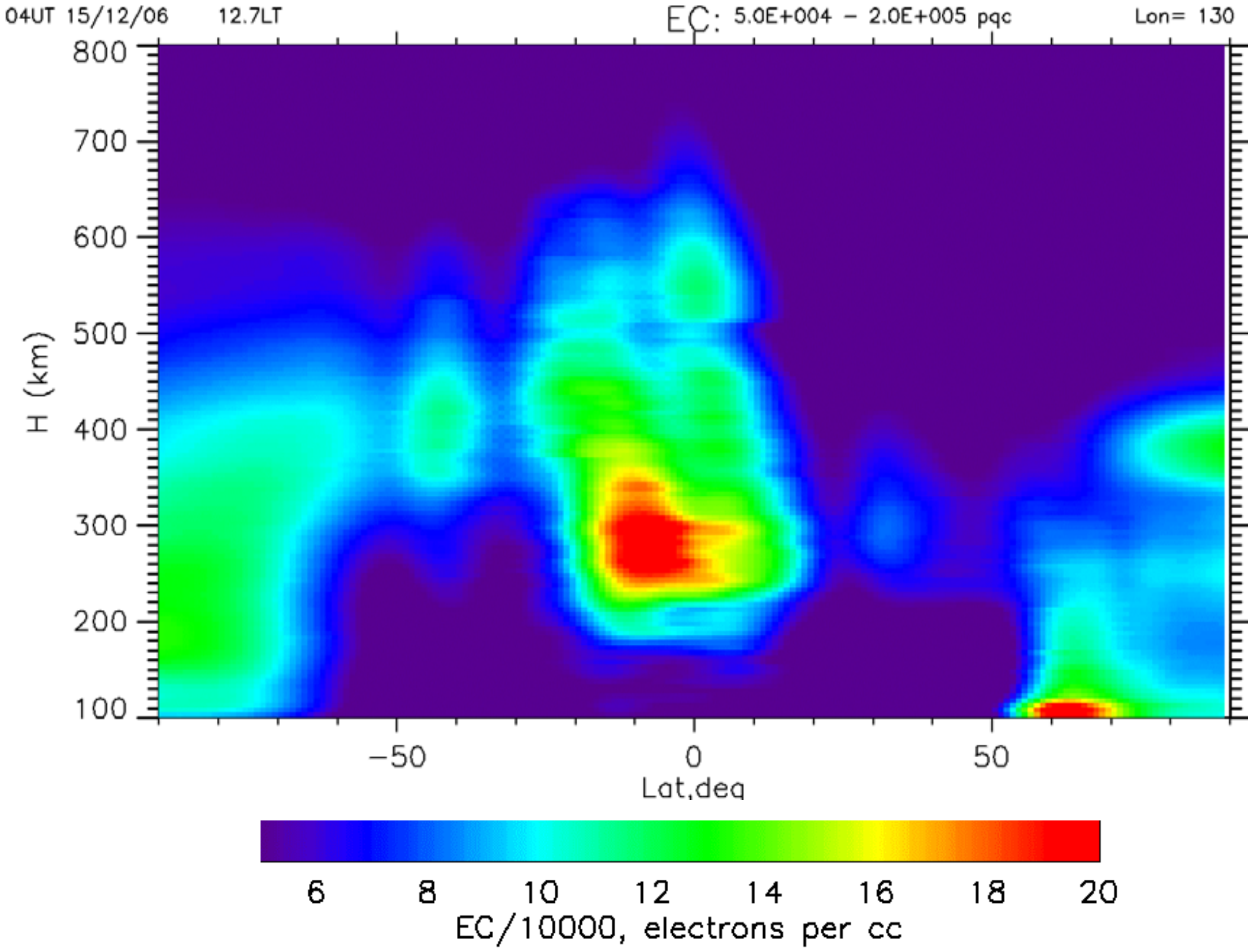

Figure 2: COSMIC/FS3 3-D tomography of electron concentration (EC) at 04 - 06 UT on 15 December 2006: a) geographic map of residual total EC (TEC) obtained by subtraction of the quiet-day TEC on December 3; b) meridional cut of EC in the range of longitudes from 130° to 135°. Geomagnetic equator is indicated by black curve. Local noon is depicted by vertical black dashed line.

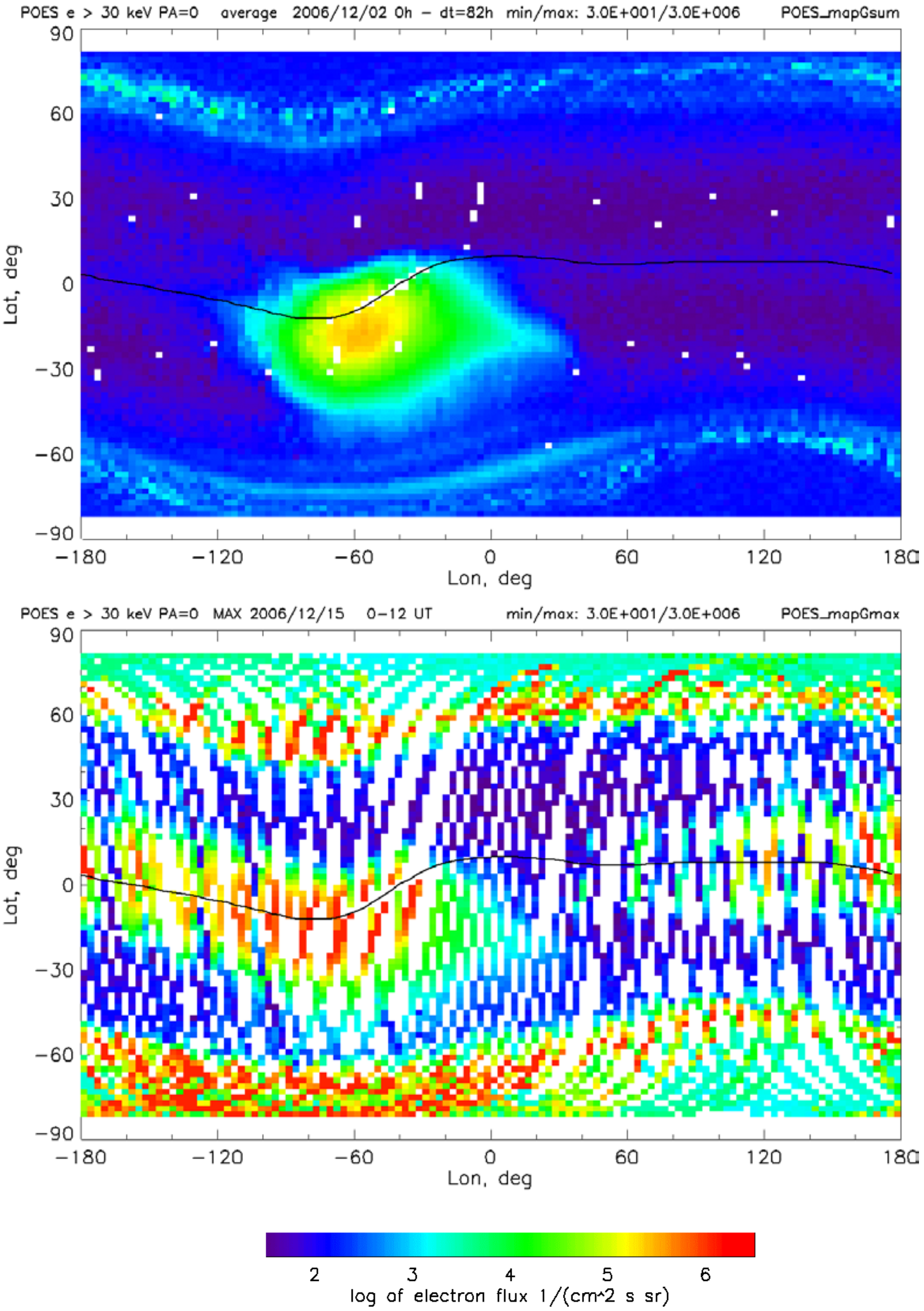

Figure 3: Geographic distribution of >30 keV electron fluxes detected by a fleet of 5 POES satellites at ~800 km altitude: (top panel) during magnetic quiet period on 2 - 4 December, 2006 and (bottom panel) during magnetic storm at 00 to 12 UT on 15 December 2006. Geomagnetic equator is indicated by black curve.

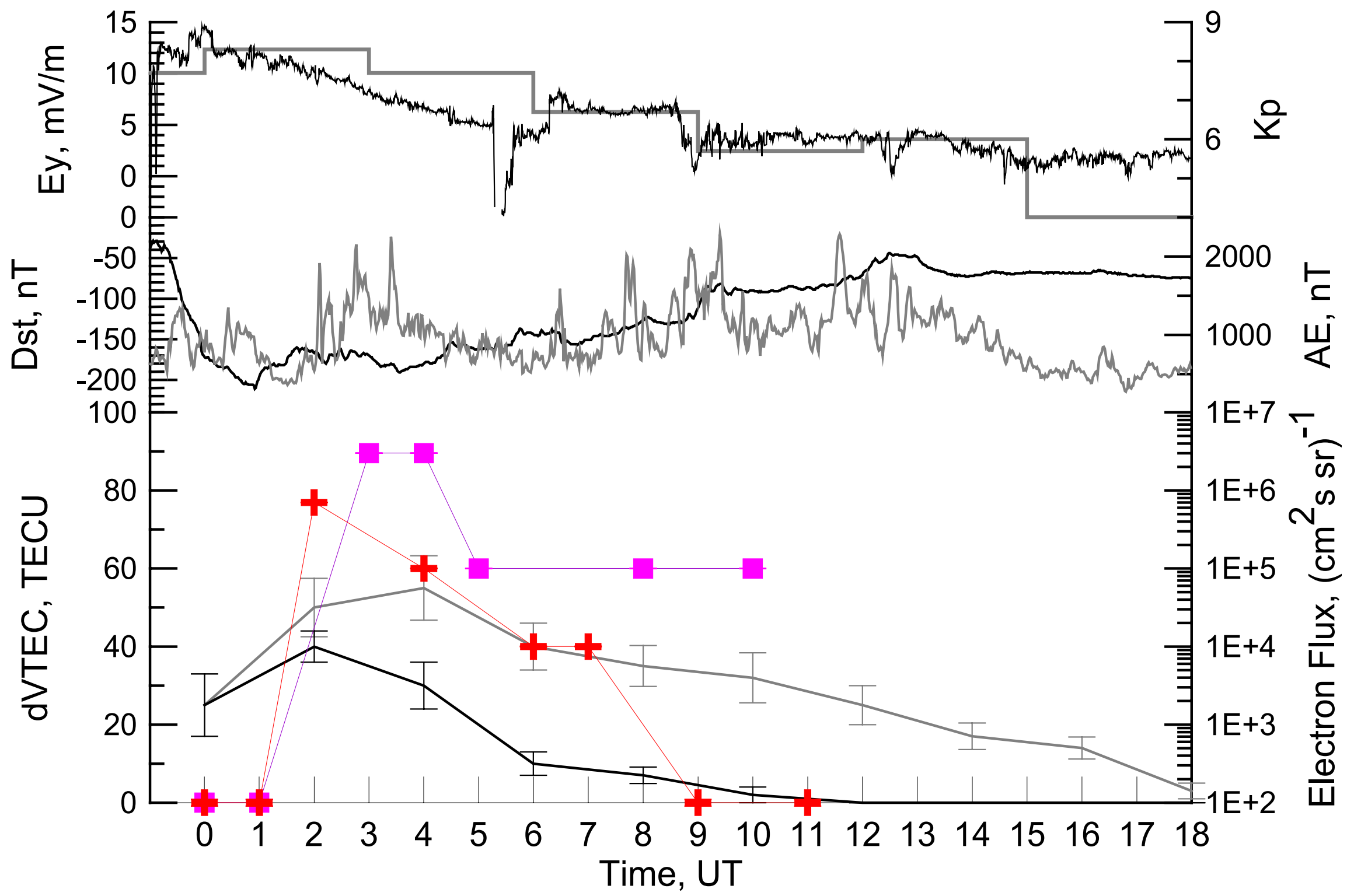

Figure 4: Variation of the interplanetary electric field *Ey*, geomagnetic indices, fluxes of >30 keV magnetospheric electrons and ionospheric residual VTEC (dVTEC) during magnetic storm from 00 to 18 UT on 15 December (from top to bottom): *Kp*-index (gray) and *E*y (black); *Dst* (black) and *AE* (gray) indices; electron fluxes at ~120° longitude (red), at ~180° (violet), maximum values of dVTEC at middle (black) and low (gray) latitudes.

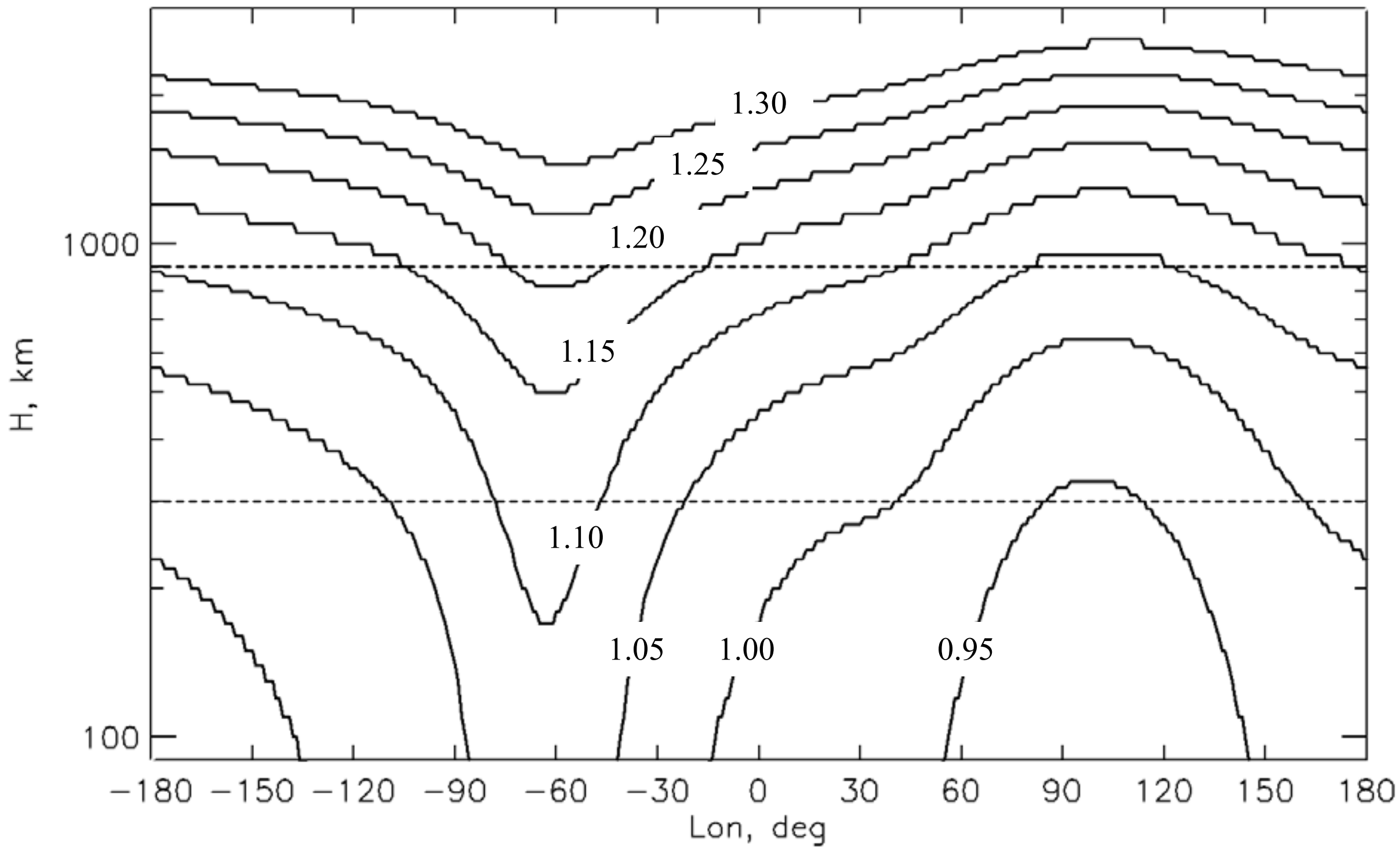

Figure 5: Longitudinal variation of the height of various drift shells at geomagnetic equator calculated from IGRF model for epoch of 2005. Quasi-trapped energetic electrons drift eastward along the drift shells and pass the highest (lowest) heights in the Indochina (SAA) region. Horizontal dashed lines indicate the heights of 300 km and 900 km.